\pdfoutput=1

\documentclass[11pt]{article}

\usepackage{acl}

\usepackage{times}
\usepackage{latexsym}

\usepackage{graphicx}
\usepackage{caption}
\usepackage{subcaption}
\usepackage{amsmath, amssymb, amsfonts}
\usepackage{svg}

\usepackage[T1]{fontenc}

\usepackage[utf8]{inputenc}

\usepackage{microtype}

\title{ColorCode: A Bayesian Approach to Augmentative and Alternative Communication with Two Buttons}

\author{Matthew Daly \\
  \texttt{mattyrdaly@gmail.com}\\}

\begin{document}
\maketitle
\begin{abstract}
Many people with severely limited muscle control can only communicate through augmentative and alternative communication (AAC) systems with a small number of buttons.
In this paper, we present the design for ColorCode, which is an AAC system with two buttons that uses  Bayesian inference to determine what the user wishes to communicate.
Our information-theoretic analysis of ColorCode simulations shows that it is efficient in extracting information from the user, even in the presence of errors, achieving nearly optimal error correction.
ColorCode is provided as open source software (\url{https://github.com/mrdaly/ColorCode}).
\end{abstract}

\section{Introduction}

People with limited muscle control, such as those affected by amyotrophic lateral sclerosis (ALS), can have trouble communicating through conventional means.
Augmentative and alternative communication (AAC) systems can help these people communicate effectively \citep{glennen1997augmentative}.
These AAC systems can range from low-tech solutions, such as pointing to messages on a piece of paper \citep{scott1998low}, to high-tech solutions, such as eye-tracking software that allows someone to select keys on a keyboard with their gaze \citep{ball2010eye}.
The more limited the muscle control, the less information that can be input into an AAC system.

In this paper, we refer to any discrete input into an AAC system as a \emph{button}, and a signal sent through a button as a \emph{click}.
Clicking a button (sometimes referred to as a \emph{switch} in AAC contexts) can take many forms, such as twitching a particular muscle or looking to the left or right.
People in the late stages of ALS may only be able to reliably click two different buttons.
AAC systems for people with such limited muscle control must efficiently extract information from a small number of buttons to allow them to communicate effectively.


An effective AAC system for users with severely limited muscle control needs to be designed to allow the individual to use a small number of buttons to choose from a large number of options (like letters on a keyboard). 
A successful design must achieve this objective while also being easy to use and resilient to errors in the user's input.
Designing a system that satisfies these properties is challenging.

There are many AAC software systems for a small number of buttons (e.g. Grid 3\footnote{Grid 3 is developed by Smartbox Assistive Technology Ltd.: \url{https://thinksmartbox.com/product/grid-3/}} and ACAT\footnote{Assistive Context-Aware Toolkit (ACAT) is provided by Intel's Open Source Technology Center: \url{https://01.org/acat/}}).
Most systems for two buttons involve a scanning keyboard where one button is used to scan through keys and the other button is used to select the chosen key \citep{colven2006switch}.
This method benefits from a simple interface, but it can take many clicks and therefore a lot of effort to communicate.

Previous research has used probabilistic reasoning and information-theoretic approaches to design effective AAC systems for few buttons (\citealp{10.1145/354401.354427}; \citealp{10.1371/journal.pone.0007481}; \citealp{7511663}).
Common themes in this research include leveraging statistical language models to improve text entry and using concepts from information theory to analyze performance.
In this work, we describe a new system whose design builds upon these existing ideas.

\begin{figure*}
    \centering
    \begin{subfigure}[b]{0.45\textwidth}
        \centering
        \includegraphics[width=\textwidth]{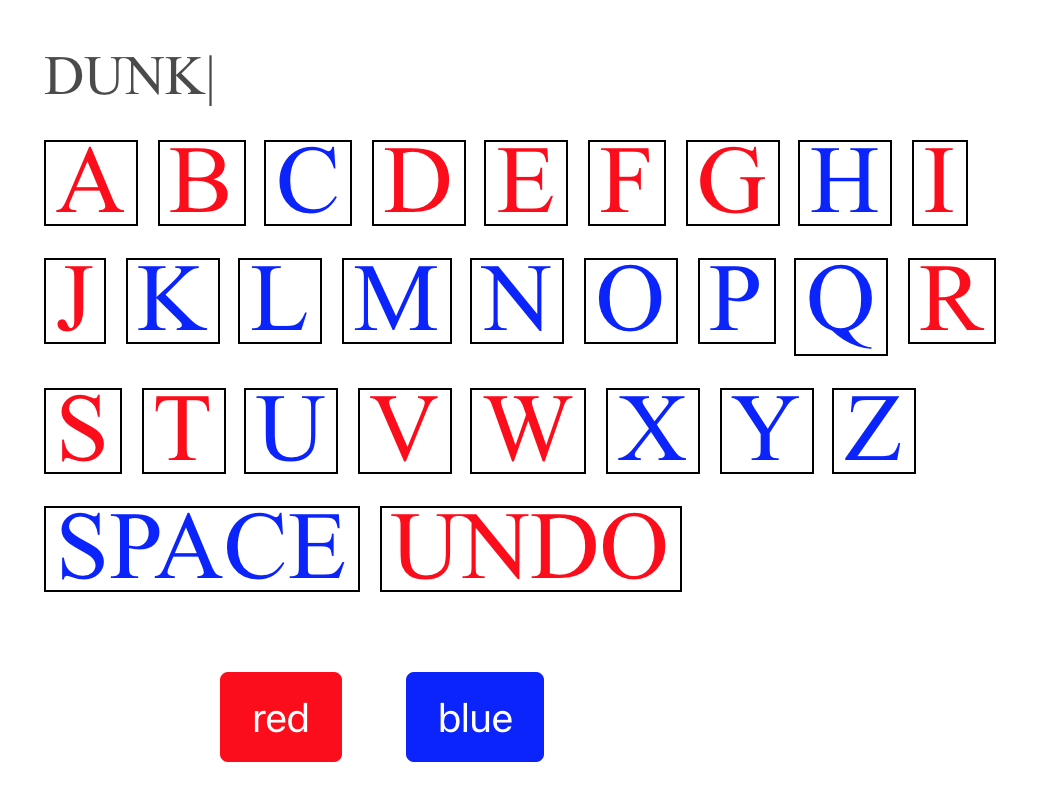}
        \caption{ColorCode interface}
        \label{fig:interface}
    \end{subfigure}
    \hfill
    \begin{subfigure}[b]{0.45\textwidth}
        \centering
        \includegraphics[width=\textwidth]{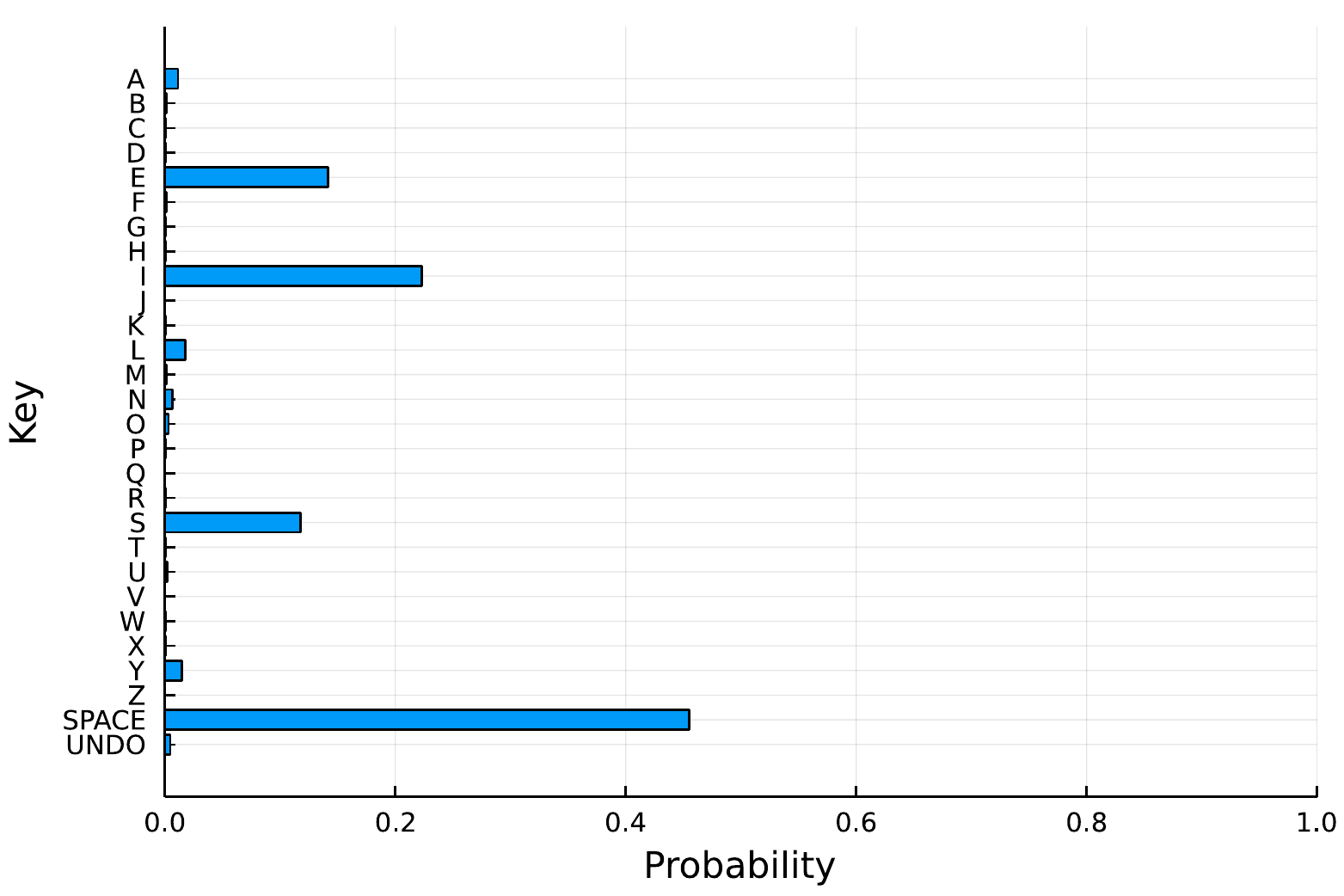}
        \caption{Belief over user's selection}
        \label{fig:belief}
    \end{subfigure}
    \caption{ColorCode system after user has begun typing.}
    \label{fig:ColorCode}
\end{figure*}

We present a new AAC system, named ColorCode, which allows users to communicate efficiently with only two buttons.
ColorCode's interface is a virtual keyboard with characters the user can select to write sentences and communicate (see Figure \ref{fig:interface}).
Each key on the keyboard is assigned one of two colors, red or blue, and each of the two buttons is associated with one of the colors.
ColorCode repeatedly assigns colors to keys while observing the colors of the buttons the user clicks.
The system uses Bayesian inference to update its belief over the user's chosen key given its observations of button clicks.
An accurate language model and the intelligent assignment of colors to keys allow ColorCode to efficiently infer what the user wishes to type.
Additionally, ColorCode adaptively corrects for errors in the user's input.

We simulate ColorCode on AAC-like text and record the average clicks per character.
We analyze the results from an information-theoretic perspective to show that ColorCode efficiently extracts information from the user.
We also simulate the input to ColorCode as a binary symmetric channel to empirically show that it is close to optimally resilient to errors in the user's input.


\section{Related Work}
\label{related_work}

This section gives a brief overview of previous research that used probabilistic reasoning and information-theoretic approaches to build AAC software systems for a small number of buttons.

\subsection{Dasher}
Dasher \citep{10.1145/354401.354427} uses the concept of arithmetic coding to allow users to efficiently type out messages.
Although the original version of Dasher requires a continuous input method like controlling a mouse pointer or joystick, other extensions allow Dasher to be controlled by a small number of discrete buttons \citep{mackay2004efficient}.

\subsection{Nomon}
AAC systems for individuals who can only click one button use the timing of the click to convey information.
Nomon \citep{10.1371/journal.pone.0007481} uses this type of input.
For each option in Nomon's interface, there is a clock with a rotating hand.
To select the option they want, the user clicks their button when the hand on the option's clock passes its noon marker.
Nomon infers the user's choice (using Bayes' rule) from the timings of their clicks, and it adaptively learns the probability distribution of the user's click timings.
The innovative design used by Nomon is very powerful and inspired much of the design of ColorCode.
However, even though Nomon is resilient to timing errors in the user's clicks, some AAC users are not able to reliably time their clicks and therefore cannot use this input method.
One of ColorCode's goals is to present a system that is as powerful as Nomon, but does not require the user to time their input.

\subsection{Shuffle Speller}
ColorCode is similar in several ways to Shuffle Speller \citep{7511663}, which is an AAC system designed for a brain-computer interface (BCI).
Shuffle Speller assigns letters to different colors associated with buttons, and the user clicks buttons to choose a color. 
The users' brain signals are interpreted through the BCI as button clicks.
Shuffle Speller uses Bayesian inference from the observed colors, and it chooses assignments of letters to colors to maximize the information it learns from observing a color.
One key difference in Shuffle Speller's design is that it accounts for asymmetry in errors across the user's inputs.
This additional complexity in modeling of input errors has the potential to improve the system's error correction.

We believe ColorCode's design is an improvement over Shuffle Speller in several ways.
First, at each color assignment, Shuffle Speller moves the letters around the screen to fixed locations in the interface associated with the colors.
In ColorCode, the letters are kept in static locations in a virtual keyboard while their colors change.
This is designed to be more user friendly, since previous research suggests dynamic keyboard layouts increase the user's cognitive load and lead to slower text entry (\citealp{lesher1998techniques}; \citealp{johansen2003language}; \citealp{pouplin:hal-01213199}).
Second, ColorCode uses adaptive ``on-the-go'' learning of the user's error rate as they use the system, but Shuffle Speller requires a calibration phase for the system to learn the distribution of the user's errors.
Finally, Shuffle Speller uses a fixed prior probability for a ``backspace'', while ColorCode incorporates evidence from the user's previous input to form a more informed prior for the ``undo'' key (see Section \ref{undo}), which is equivalent to Shuffle Speller's backspace.

\section{Method}
\label{methods}

\begin{figure}
    \centering
    \includegraphics[width=0.4\textwidth]{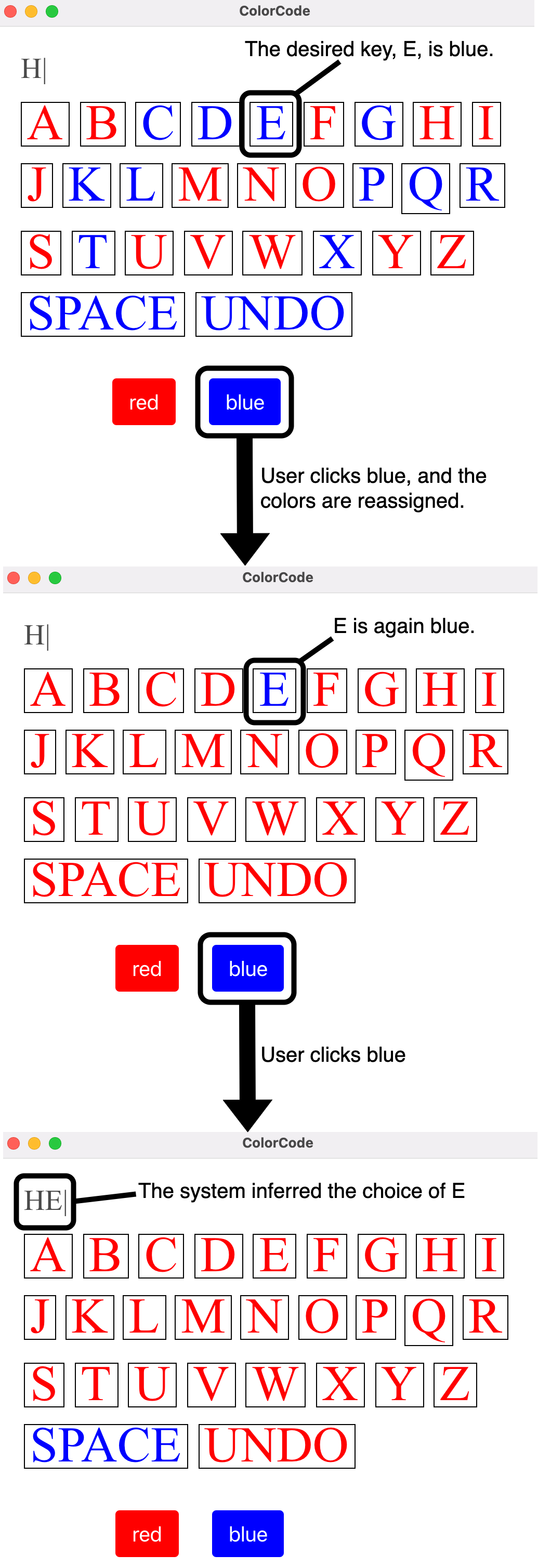}
    \caption{Diagram demonstrating process of selecting the letter E in ColorCode.}
    \label{fig:diagram}
\end{figure}

To type a message in ColorCode, the user first identifies the color of the key they wish to select (e.g. the letter A) and clicks the button for that color.
After the user clicks a button, the system reassigns colors to all of the keys on the keyboard.
The user repeats this process until the system selects the key they wanted and types the corresponding character in the display (or deletes a character if key was ``undo'').
ColorCode also plays an audible ``click'' sound to notify the user when a key is selected.
See Figure \ref{fig:diagram} for a diagram demonstrating this process.

ColorCode maintains a belief over the user's desired key and uses Bayes' rule to update its belief after observing the user's button click.
The belief is a probability distribution over the possible keys (see Figure \ref{fig:belief}).
When the probability of a particular key reaches a certain threshold, the system selects that key.
This probability threshold is set to $0.95$ in ColorCode. 
If $P(k)$ is the probability the user wishes to select key $k$, and $c$ is the color of the button the user clicks, we can compute the belief update using Bayes' rule with:
\[P(k \mid c) = \frac{P(c \mid k)P(k)}{\sum_{k'} P(c \mid k')P(k')}\]
In this update, the two key components are the probability distributions $P(k)$ and $P(c \mid k)$, which are known as the \emph{prior} and the \emph{likelihood}, respectively.

\subsection{Prior}
The prior is a probability distribution representing the previous belief over the user's key selection before the user clicks a button.
When the user has already started clicking buttons to select a key, the prior is simply the output of the previous belief update.
However, when beginning a key selection with no previous button clicks, we can define the prior based on any knowledge we have about which key the user will select.
If we had no prior knowledge, the prior would be a uniform distribution over all the possible keys.

ColorCode uses a language model that uses the context of what the user previously typed to estimate the probability distribution over which character comes next.
The 12-gram character language model\footnote{\url{https://imagineville.org/software/lm/dec19_char/}} used by ColorCode was trained on billions of words of AAC-like text, providing a  well-informed prior that allows ColorCode to infer the user's key selection in a few button clicks.

\subsection{Likelihood}
\label{likelihood}
The likelihood, $P(c \mid k)$, is the probability that ColorCode would observe the color $c$ from the user's click, given that they want to select the key $k$.
If ColorCode always observed the user click the correct color button for the key they wished to select, then $P(c \mid k)$ would be $1$ if $c$ was the color of key $k$ and $0$ otherwise.
However, it is possible that the user could make a mistake or the button's sensor could be noisy, and the system could observe a click for the wrong color.
To account for these possible errors, we define the likelihood as the probability that the system observes the user click the correct color given their desired key and its color.

The system does not know the probability of a click error, so it must estimate it in some way.
We assume the distribution of click errors is a stationary binary distribution with the parameter $\theta$, which is the probability that the click is correct.
ColorCode uses Bayesian learning to learn $\theta$ from observing correct and incorrect clicks from the user.
Since the beta distribution is the conjugate prior of the binary distribution, we have:
\[P(\theta) = \text{Beta}(\theta \mid \alpha,\beta)\]
where $\alpha$ and $\beta$ are parameters of the beta distribution.
Each time the system observes the user correctly clicking a color button given their key, it increments $\alpha$ by $1$, and it increments $\beta$ when it observes the user clicking the incorrect button.

The parameters $\alpha$ and $\beta$ can both be initialized to $1$ to represent a uniform distribution over $\theta$, but ColorCode starts with pseudocounts of $\alpha=9$ and $\beta=1$ to encode the belief that the probability of click errors is low.
When computing the belief update, ColorCode uses the mean of the beta distribution:
\[ P(c \mid k) = \begin{cases} 
      \frac{\alpha}{\alpha + \beta} & \text{if } c \text{ is the color of } k \\
      \frac{\beta}{\alpha + \beta} & \text{if } c \text{ is not the color of } k \\
   \end{cases}\]

It is not obvious how we can count when a user's click is correct or incorrect.
ColorCode obtains these counts by keeping track of the user's clicks and the color assignments to keys, and then once a key is selected, it goes back and updates $\alpha$ and $\beta$ assuming the selected key was the correct one.

When this likelihood is applied to the belief update, it corrects for the probability of error.
If click errors are more likely, then the update increases (or decreases) the probability of keys by a smaller factor, therefore requiring more clicks to select a letter.

\subsection{Undo}
\label{undo}

We include an ``undo'' key in ColorCode's interface to allow the user to indicate that the system incorrectly inferred the previous key selection.
Many AAC systems have a similar option, often referred to as ``backspace'' or ``delete''.
Many probabilistic AAC systems incorporate an undo key into the prior by fixing its probability to a predefined constant, such as $0.05$, and then normalizing the rest of the keys' probabilities (e.g. \citealp{6287966}; \citealp{7511663}). 
\citet{10.1145/2513383.2513453} introduces an algorithm that keeps track of probabilities for alternative inferences the system could have made, and then uses these probabilities to inform the prior probability of a backspace key.

In ColorCode, we set the undo probability to the probability that the previous key selection was wrong.\footnote{During the first key selection, the undo probability is set to $0$, since there are no key selections to undo.}
We know this probability from the belief the system had when it selected the previous key. 
When starting a new key selection, we have:
\[P(k_t = \text{UNDO}) = 1 - P(k_{t-1}=K)\]
where $t$ is the current time step in terms of  belief updates, and $K$ is the key the system previously selected.

When the user selects the undo key, the last character in the output string is removed.
ColorCode then assumes that the selection of the removed character was incorrect, and the user still wishes to select the key that they originally intended to select.
With this assumption, the system resets the prior to the belief it had at the time of the incorrect selection.
However, it changes the probability of the assumed incorrect character to be the probability that the undo selection was wrong.
After the system selected undo at step $t-1$, we have:
\[P(k_t = K) = 1 - P(k_{t-1}=\text{UNDO})\]
where $K$ is the key the system is assumed to have incorrectly selected before the undo.
The probabilities of the rest of the keys are then normalized.

Additionally, when the user selects undo, ColorCode undoes the error rate learning it did on the previous key selection, since it can no longer assume that the selection was correct.

\subsection{Color Assignment}
An important aspect of ColorCode is the assignment of colors to keys at each step.
This assignment determines what information the system learns when it observes a button click.
In the degenerate case where all keys are assigned the same color, no information can be learned from a user's click.
The goal of choosing a color assignment is to learn as much information as possible about the user's desired key selection.

We can use the entropy of observing a color as a heuristic to measure the effectiveness of a color assignment.
The entropy, which can be thought of as the expected information content received from observing a user's color click, is defined as:
\[-\sum_c P(c) \log P(c) \]
where $P(c)$ is the probability of observing a click of color $c$ based on our current belief.

Computing the entropy for every possible color assignment and choosing the maximum is intractable, but we can also maximize entropy by maximizing the uniformity of the probability distribution, $P(c)$ \citep{10.5555/971143}. 
Choosing a color assignment that makes $P(c)$ as close to equiprobable as possible is equivalent to the partition problem, which is NP-complete.
However, there are approximate algorithms for the partition problem that run in polynomial time, and ColorCode uses the simple greedy heuristic to approximate a solution \citep{10.5555/1625855.1625890}. 

We also tried assigning colors using Huffman coding, similar to \citet{ROARK20131212}.
While our simulations showed the Huffman coding approach performed slightly better than the partition approach with no click error, the partition approach performed better in the presence of errors.
For this reason, ColorCode uses the partition approach for color assignments.
A possible alternative would be to use Huffman coding when the system estimates an error rate below a certain threshold, and then use the partition approach if the error rate is above that threshold.

\section{Results and Analysis}
\label{results}
The effectiveness of ColorCode can be measured by the average number of clicks it takes the user to select a character.
Low clicks per character (cpc) indicate that a system is efficient in its ability to extract information from the user.

Another important metric used to evaluate AAC systems is the text entry rate (TER).
Measuring the text entry rate of a system requires a user study, which has not yet been performed using ColorCode (see Section \ref{future_work}).

We simulated ColorCode on a test set of AAC-like text, presented in \citet{vertanen_aac_lm}, to calculate the average cpc.
The simulator uses the undo key to correct any incorrect key selections, and these extra clicks are counted toward the cpc.
Through these simulations, ColorCode achieved an average of $2.07$ cpc.

We can analyze this result from an information-theoretic perspective by considering the theoretical lower bound on cpc given the language model ColorCode uses.
We can view the color clicks as binary symbols (bits) in a variable-length encoding of the characters the user wishes to type.
The source coding theorem for symbol codes states that the entropy of a character distribution is the lower bound on the expected number of bits required to encode a character \citep{10.5555/971143}.
However, the system uses a language model for the encoding because it does not know the true character distribution. 
So instead, the cross-entropy between the true distribution and the model distribution can be used as a lower bound on the expected number of bits per character in a coding scheme that uses the model distribution to encode characters that come from the true distribution \citep{brown1992estimate}.
The cross-entropy is defined as:
\[-\mathbb{E}_{x \sim P(x)} [ \log M(x \mid x_{-1}, x_{-2}, \dots)]\]
where $P(x)$ is the true distribution of characters and $M(x \mid x_{-1}, x_{-2}, \dots)$ is the language model's probability of a character given its context.
To calculate the cross-entropy empirically, we estimate the expectation over the true distribution by averaging over the AAC test set used to evaluate ColorCode. 
Using this approach with the language model that ColorCode uses, we calculate the cross-entropy to be $1.73$ bits.
This means that ColorCode achieves $2.07$ cpc when the lower bound given its language model is $1.73$ cpc.

\subsection{Error Correction}
To evaluate ColorCode's resilience to click errors, we ran simulations on the test set with a parameter $f$, which defined the probability that the simulator would randomly click the incorrect color for the desired key.

Let us consider the user's clicks as bits being communicated over a noisy channel, specifically a binary symmetric channel (BSC).
The BSC has a probability $f$ of a bit flip (the color is incorrect) and a probability of $1-f$ of a correct bit transmission (the color is correct).
Error-correcting codes can be used to communicate over noisy channels by sending more bits for redundancy.
Recall that ColorCode learns the click error rate and then requires more clicks from the user to compensate for more errors.
Let us think of this mechanism in ColorCode as an error-correcting code.

We can evaluate an error-correcting code by its \emph{information rate}, which is the ratio of information bits communicated to the total number of bits sent over the channel.
The total number of bits includes both the information bits and the redundant bits which are sent to correct any errors.
We can measure the information rate of ColorCode by using simulations on the test set.
We define the number of information bits as the number of clicks required during a simulation with no error rate.
Then we define the total number of bits as the total number of clicks the simulation requires when given an error rate $f$.
With this, we can calculate the information rate of ColorCode for a given error rate.

According to the noisy-channel coding theorem, the error rate of a noisy channel can be corrected to an arbitrarily small resulting error \citep{10.5555/971143}. 
Additionally, any error-correcting code that can achieve this has a maximum information rate equal to the \emph{channel capacity}, $C$, of the channel.
The channel capacity of a BSC with error rate $f$ is
\[C = 1 - h_2(f)\]
where $h_2$ is the binary entropy function.

\begin{figure}
    \centering
    \includegraphics[width=0.4\textwidth]{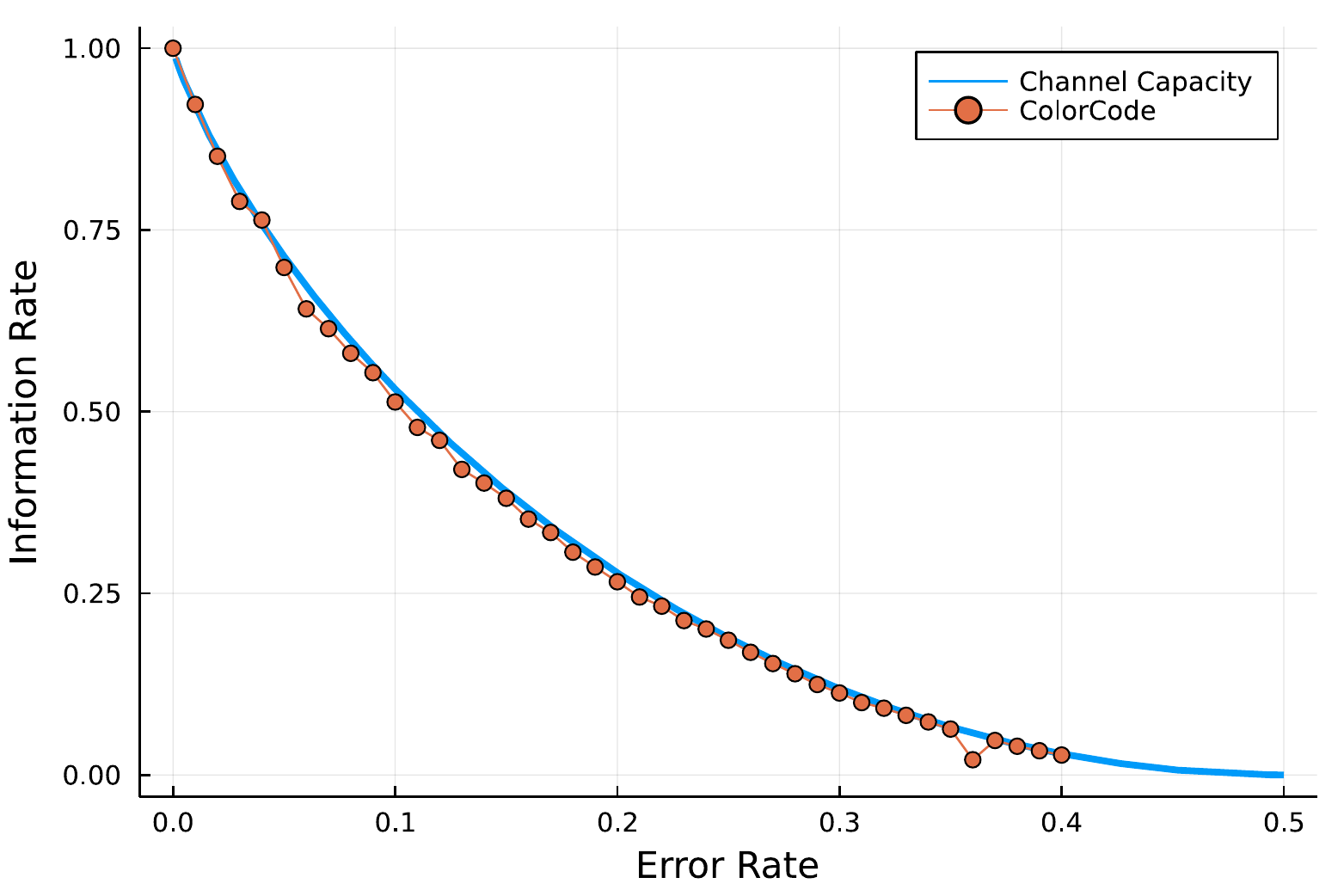}
    \caption{Plot comparing the information rate of ColorCode to the channel capacity of a binary symmetric channel.}
    \label{fig:info_rate}
\end{figure}

We plot the information rates from our simulations in Figure \ref{fig:info_rate}, along with the optimal information rates of a BSC, the channel capacity.
These empirical results show that ColorCode's error correction is nearly optimal when we model the errors with a BSC.

\section{Conclusions}
This paper presents the design of ColorCode, a new and powerful AAC system for two buttons.
ColorCode combines a powerful language model, Bayesian learning of click errors, an informed undo operation, and intelligent color assignments into a Bayesian belief framework that uses a simple interface to efficiently extract information from the user.
Our results demonstrate this efficiency by showing ColorCode requires an average of only $2.07$ clicks to select a character, which is within one half of a click of the theoretical lower bound which is $1.73$ clicks.
ColorCode remains efficient in extracting information when there are errors in the input, and our results show that ColorCode handles errors with nearly optimal efficiency.
These results show that ColorCode's design has the potential to help people who cannot communicate easily.

\section{Future Work}
\label{future_work}
Further development on ColorCode can make it viable for real-world use as an AAC system.
Several improvements to the design could increase performance by making it easier for users to convey information.
One improvement would be to extend ColorCode to optionally use more than two colors, which would help if the user has more control and can click more than two buttons.
We could also improve the design with word predictions or other possible methods that would leverage the powerful language model to let the user select multiple characters with one click.

Additionally, conducting a user study with ColorCode is a vital next step in its development.
Testing ColorCode with real users and real input devices is essential to evaluating its text entry rate, interface usability, and error correction.


\section*{Acknowledgements}
Special thanks to Mykel Kochenderfer and Keith Vertanen.

\bibliographystyle{acl_natbib}
\bibliography{custom}

\end{document}